\documentclass[prd,aps,twocolumn,nofootinbib,preprintnumbers,superscriptaddress,preprintnumbers,balancelastpage,longbibliography]{revtex4-2}
\pdfoutput=1
\usepackage{amsmath,mathtools,physics,xfrac}
\usepackage{graphicx}
\usepackage{afterpage}
\usepackage{float}
\usepackage{subfigure}
\usepackage{rotating}
\usepackage{multirow}
\usepackage{tabularx}
\usepackage{booktabs}
\usepackage{fancyhdr}
\usepackage[utf8]{inputenc}
\usepackage{theorem}
\usepackage{moreverb}
\usepackage{euscript}
\usepackage{psfrag}
\usepackage{slashed}
\usepackage{mathtools}
\usepackage{makecell}
\usepackage{adjustbox}
\usepackage{dcolumn}
\usepackage{bm}
\usepackage[dvipsnames]{xcolor}
\usepackage{graphics}
\usepackage{hyperref}
\usepackage[shortlabels]{enumitem}
\usepackage{cancel}

\definecolor{nice}{rgb}{0.103, 0.148, 0.255}

\hypersetup{
     colorlinks   = true,
     citecolor    = {MidnightBlue},
     urlcolor     = {MidnightBlue},
     linkcolor    = {MidnightBlue}
}

\begin{document}

\preprint{UCI-HEP-TR-2024-19}

\title{Strong Constraints on Dark Photon and Scalar Dark Matter Decay \\ from INTEGRAL and AMS-02 data}

\author{Thong T.Q. Nguyen}
\thanks{\href{mailto:thong.nguyen@fysik.su.se}{thong.nguyen@fysik.su.se}, \href{https://orcid.org/0000-0002-8460-0219}{ORCID: 0000-0002-8460-0219}}
\affiliation{Stockholm University and The Oskar Klein Centre for Cosmoparticle Physics,  Alba Nova, 10691 Stockholm, Sweden}

\author{Isabelle John}
\thanks{\href{mailto:isabelle.john@unito.it}{isabelle.john@unito.it}, \href{https://orcid.org/0000-0003-2550-7038}{ORCID: 0000-0003-2550-7038}}
\affiliation{Dipartimento di Fisica, Universit\`a degli Studi di Torino, via P.\ Giuria, 1 10125 Torino, Italy}
\affiliation{INFN -- Istituto Nazionale di Fisica Nucleare, Sezione di Torino, via P.\ Giuria 1, 10125 Torino, Italy}
\affiliation{Stockholm University and The Oskar Klein Centre for Cosmoparticle Physics,  Alba Nova, 10691 Stockholm, Sweden}

\author{Tim Linden}
\thanks{\href{mailto:linden@fysik.su.se}{linden@fysik.su.se}, \href{https://orcid.org/0000-0001-9888-0971}{ORCID: 0000-0001-9888-0971}}
\affiliation{Stockholm University and The Oskar Klein Centre for Cosmoparticle Physics,  Alba Nova, 10691 Stockholm, Sweden}

\author{Tim M.P. Tait}
\thanks{\href{mailto:ttait@uci.edu}{ttait@uci.edu}, \href{https://orcid.org/0000-0003-3002-6909}{ORCID: 0000-0003-3002-6909}}
\affiliation{Department of Physics and Astronomy, University of California, Irvine, CA 92697 USA}

\begin{abstract}
\noindent We investigate the decay of bosonic dark matter with masses between 1~MeV and 2~TeV into Standard Model final states. We specifically focus on dark photons that kinetically mix with the Standard Model, as well as scalar dark matter models that have Yukawa couplings with the Standard Model. Using INTEGRAL and AMS-02 data, we constrain the dark matter decay lifetime into final states that include photons or positrons, setting strong constraints on the dark matter lifetime that reach 10$^{25}$~s for dark matter below 10~GeV and up to 10$^{29}$~s for dark matter above 10~GeV.
\end{abstract}

\maketitle

\section{Introduction}
\label{sec:intro}

The possible interactions of dark matter with the Standard Model (SM) stand among the most perplexing puzzles in physics~\cite{Bertone:2004pz, Bertone:2016nfn}. To determine the properties of dark matter, physicists and astronomers have developed techniques that depend on the ``indirect" observations of stable SM particles that may be produced by dark matter interactions~\cite{Bertone:2018krk, Slatyer:2017sev, Safdi:2022xkm}. One compelling strategy is to search for the possible late-time decays of long-lived dark matter particles to stable SM final states such as photons, positrons, antiprotons, and neutrinos. Many studies of dark matter decay, spanning from sub-GeV energies~\cite{Cirelli:2023tnx, Cirelli:2020bpc, DelaTorreLuque:2025zjt} through the TeV-scale~\cite{Liu:2020wqz, Qin:2023kkk, Jin:2013nta, Dutta:2022wuc, Ghosh:2020ipv}, and potentially up to the Planck scale~\cite{Blanco:2018esa, Das:2024bed, Murase:2012xs, Song:2024vdc, Song:2023xdk, Das:2023wtk, Acharyya:2023ptu, Bauer:2020jay, Rinchiuso:2018ajn, Cohen:2016uyg, Baldes:2020hwx}, have exploited these multi-messenger indirect detection signals to set strong constraints on the dark matter lifetime.

In many cases, however, these studies consider a generic model where the dark matter particle decays into a single pair of SM particles, for example $e^+e^-$ or $b\bar{b}$. This ignores the fact that the majority of realistic dark matter models include mediators that simultaneously couple to multiple SM particles (for example, based on their charge or mass), with ratios that often change as a function of the dark matter mass~\cite{Cirelli:2016rnw, Cirelli:2024ssz}. While such an approach improves the ability to compare results between different experiments, it makes it difficult to understand the resulting constraints on dark matter model building.

Particle-antiparticle final states can be produced via the decays of bosonic dark matter particles. There are several well-known and self-consistent models that are worth studying in detail~\cite{Pospelov:2008jd, Pospelov:2008jk}. In this paper, we consider two such classes of models: (1) vector dark photon dark matter~\cite{Fabbrichesi:2020wbt, Rizzo:2018joy, Rizzo:2018ntg, Hebecker:2023qwl, Fayet:1980ad, Fayet:1990wx, Pospelov:2008jk, Servant:2002aq, Fradette:2014sza, Kitajima:2024jfl, Cyncynates:2024yxm} that couples to the SM via kinetic mixing with neutral gauge bosons, and (2) scalar dark matter~\cite{Pospelov:2008jk} that has Yukawa-like couplings with the SM fermions. In their simplest realizations, these models determine the exact branching ratios for all decay channels as a function of the dark matter mass. Focusing on these models allows us to self-consistently consider a wide range of potential dark matter masses, spanning from two electron masses up to the Planck mass.

In this paper, we focus on the mass range below 2~TeV, where sensitive observations exist. First, we calculate the complete branching ratios for dark photon and scalar dark matter decays, and then use x-ray observations from the INTErnational Gamma-Ray Astrophysics Laboratory (INTEGRAL) satellite to constrain dark matter masses below 10~GeV, along with positron observations from the Alpha-Magnetic Spectrometer (AMS-02) to constrain heavier dark matter particles. We find no evidence for a dark matter signal, and rule out dark matter decay lifetimes below 10$^{25}$s for MeV-scale dark matter, and 10$^{29}$s for GeV-scale dark matter. We discuss the implications of these limits compared to previous work which focused on individual SM final states, and discuss the impact of future x-ray and cosmic-ray observations on dark photon and scalar particles.

This paper is organized as follows. In Sec.~\ref{sec:model}, we discuss the Lagrangian for dark photon and scalar dark matter particles and calculate the decay rates for each channel. In Sec.~\ref{sec:data}, we discuss the indirect signals for photon and positron final states, for which we provide detailed formulas of photon spectra in Appendix~\ref{appen:photon}. In Sec.~\ref{sec:result}, we show our results, which place strong constraints on the dark matter decay lifetime. We also compare our results with single decay models in Sec.~\ref{sect:single}. Finally, we summarize our study in Sec.~\ref{sec:summary} and discuss the applications of our work to other dark matter models. We utilize natural units ($c=\hbar=1$) throughout this manuscript.

\section{Dark Matter Models and Decays}
\label{sec:model}
In this section, we discuss the two dark matter models in the following order: the dark photon dark matter model which interacts with the SM via kinetic mixing with SM neutral gauge bosons, and then scalar dark matter that mixes with the SM Higgs boson. We then calculate the decay width to each SM channel, providing branching ratios to SM final states that depend on the dark matter mass. 

\subsection{Dark photon dark matter}
\label{subsec:darkphoton}

The dark photon particle is a consequence of a new $U(1)$ symmetry~\cite{Fabbrichesi:2020wbt, Rizzo:2018joy, Rizzo:2018ntg, Hebecker:2023qwl, Fayet:1980ad, Fayet:1990wx, Carenza:2025uwx} that is added to the SM Lagrangian. After symmetry breaking, the Lagrangian is:
\begin{equation}
    \mathcal{L}\supset -\frac{1}{4}F^{\prime}_{\mu\nu}F^{\prime \mu\nu}-\frac{1}{2}m_{A^{\prime}}^{2}A^{\prime}_{\mu}A^{\prime\mu}-\frac{\epsilon}{2\cos\theta_{w}}F^{\prime}_{\mu\nu}B^{\mu\nu},
    \label{eq:Lagrangian_DP}
\end{equation}
where $A^{\prime}$ is the dark photon field, with its mass $m_{A^{\prime}}$, $F_{\mu\nu}^\prime$ is the dark photon field strength tensor and %is the dark matter mass and
$\epsilon$ is the kinetic coupling. The $B_{\mu\nu}$ is the SM neutral gauge boson field strength tensor, with the weak interaction mixing angle $\theta_{w}$. Qualitatively, for dark matter masses that approach the weak-scale, the dark photon model becomes similar to a $Z^{\prime}$ gauge boson model that includes kinetic mixing with neutral SM vector bosons~\cite{Carena:2004xs, Langacker:2008yv}. This new vector boson is well motivated by other phenomenological studies~\cite{Caputo:2021eaa} as a new mediator that connects the dark sector with the SM~\cite{Jaeckel:2023huy, Fayet:1981rp, Nguyen:2023ugx, Tran:2023lzv, Linden:2024uph, Nguyen:2022zwb, Krnjaic:2022wor, Fitzpatrick:2021cij, Fitzpatrick:2020vba, Giovanetti:2021izc, Emken:2024nox}, and potentially provides a stable dark matter candidate~\cite{Linden:2024fby, Arias:2012az, Redondo:2008ec, Bogorad:2023wzn, Graham:2015rva, DelaTorreLuque:2024zsr, DelaTorreLuque:2023huu, DelaTorreLuque:2023nhh, McDermott:2019lch, Catena:2022fnk, Witte:2020rvb, Caputo:2020bdy, Krnjaic:2023nxe, An:2014twa, Nguyen:2025eva}.

\begin{figure}[t]
\centering
\includegraphics[width=0.8\columnwidth]{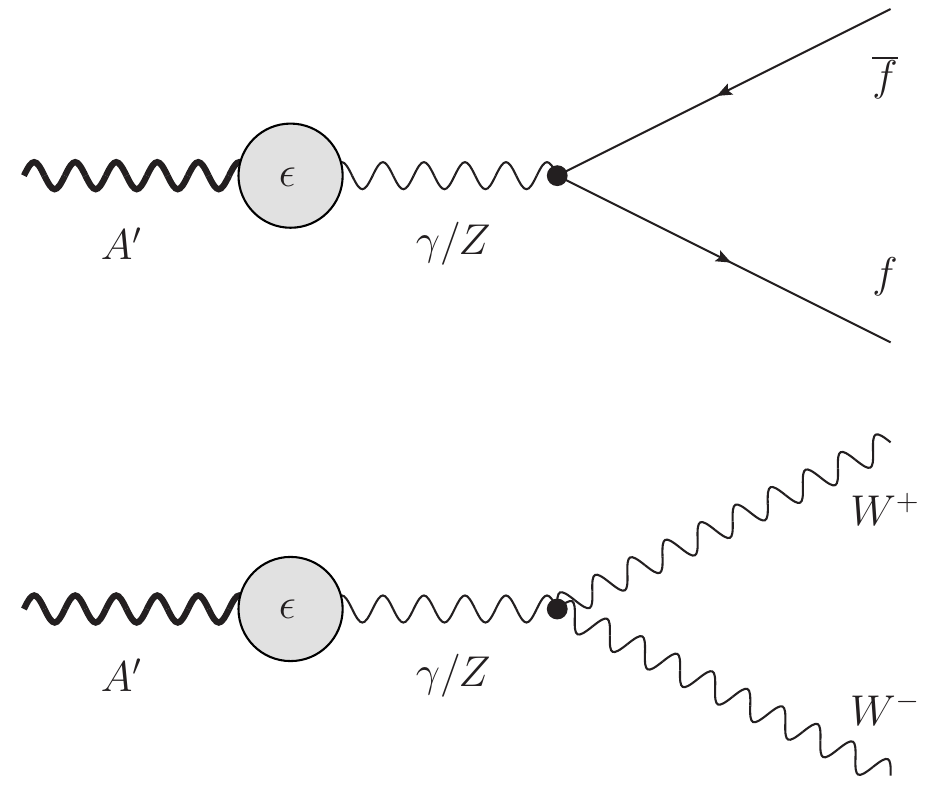}
\caption{Feynman diagrams for dark photon dark matter decay to SM particles through mixing with SM neutral gauge bosons. From 1~MeV to 161~GeV, the dark photon mainly decays into fermion pairs with EW couplings that are proportional to the fermion charges. Dark photon with masses above 161~GeV can decay to $W$-boson pair through $AW^{+}W^{-}$ and $ZW^{+}W^{-}$ couplings.}
\label{fig:DPdecay}
\end{figure}

In scenarios where the dark photon is the dark matter particle, it can directly decay to SM particles through a non-zero kinetic mixing with the SM neutral gauge bosons. In recent work, we have carefully studied and constrained dark photon decay in the sub-MeV range, where their dominant decay mode produces 3~$\gamma$-ray final states~\cite{Linden:2024fby}. In this study, we instead focus on dark photon dark matter above 1~MeV, which kinematically allows for dark photon decay into SM fermions and $W$-bosons.

\begin{figure}[t]
\centering
\includegraphics[width=1\columnwidth]{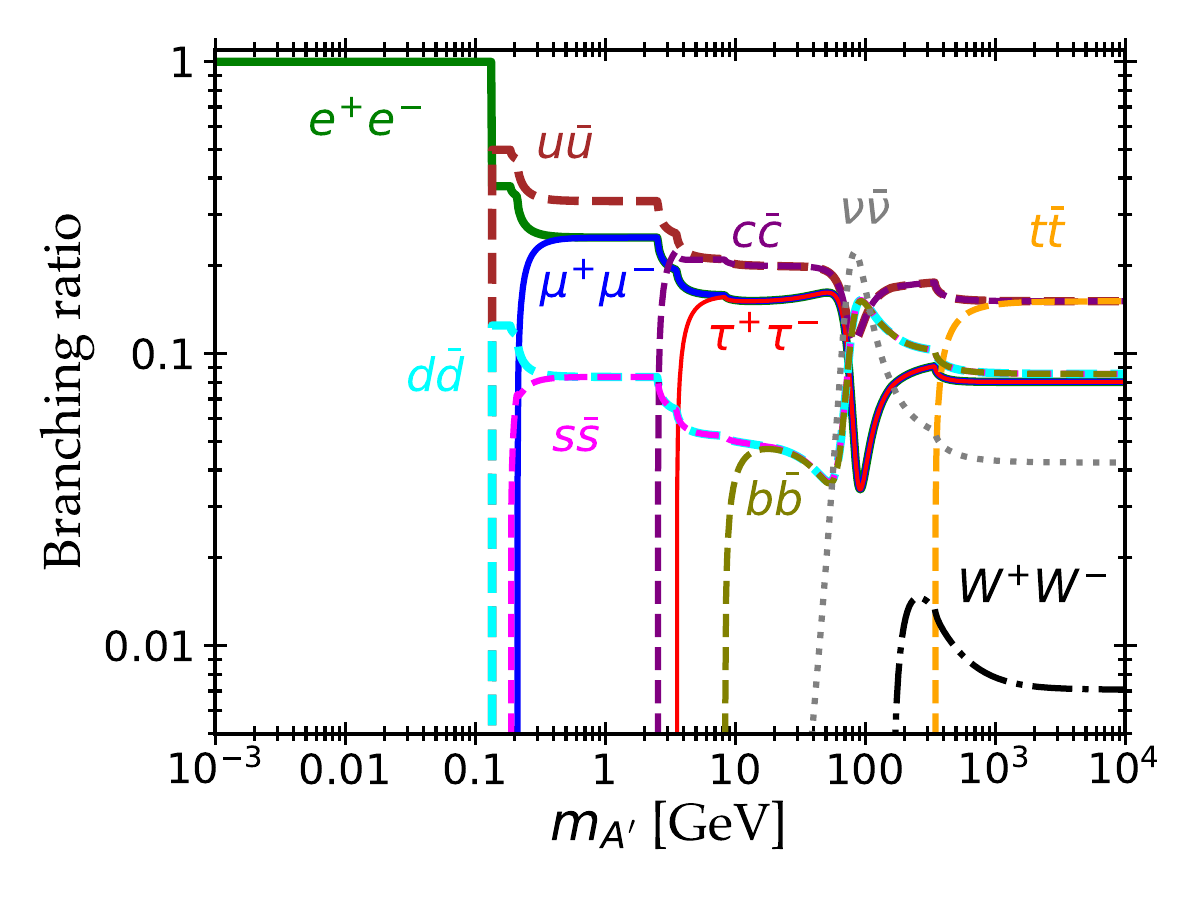}
\caption{Branching ratio of all Standard Model channels for dark photon decay. The solid lines are for charged leptons, while the dashed lines are for quark final states. The dash-dotted line is for $W$-pair production, while the dotted line is for all neutrino final state. These results are also applied to vector boson decays with EW-like couplings.}
\label{fig:BRDP}
\end{figure}

In models where the dark photon mass lies well below the $Z$-boson mass, the dark photon decay rate to fermions is typically approximated following the equation given in Ref.~\cite{Fabbrichesi:2020wbt} as:

\begin{equation}
    \Gamma\underset{m_{A^{\prime}}\ll m_{Z}}{(A^{\prime}\to \bar{f}f)}=\frac{\alpha N_{c}Q_{f}^{2}\epsilon^{2}}{3}m_{A^{\prime}}\sqrt{1-\frac{4m_{f}^{2}}{m_{A^{\prime}}^{2}}}\Big{(}1+\frac{2m_{f}^{2}}{m_{A^{\prime}}^{2}}\Big{)},
    \label{eq:Gamma_Aff}
\end{equation}
where $\alpha=1/137$ is the fine structure constant, $Q_f$ is the electric charge of the SM particle, $m_f$ is the mass of the SM fermion, and $N_{c}$ represents the number of colors for each Standard Model fermion, with $N_{c}=1$ for charged leptons, and $N_{c}=3$ for charged quarks. However, as the dark photon mass approaches the $Z$-boson mass, contributions from kinetic mixing with the $Z$ become dominant. In this work, we calculate the full decay width to SM fermions, including electroweak (EW) interactions as:
\begin{align}
    \Gamma(A^{\prime}\to \bar{f}f)&=\frac{\alpha N_{c}\epsilon^{2}}{6 c_{w}^{4}m_{A^{\prime}}[(m_{A^{\prime}}^{2}-m_{Z}^{2})^{2}+m_{Z}^{2}\Gamma_{Z}^{2}]}\\
    &\times(F_{1}+F_{2}+F_{3})\sqrt{1-\frac{4m_{f}^{2}}{m_{A^{\prime}}^{2}}}\nonumber,
\end{align}
where the functions $F_{1,2,3}$ are defined as:
\begin{align}
    F_{1}&=I_{3,f}^{2}m_{A^{\prime}}^{4}(m_{A^{\prime}}^{2}-m_{f}^{2}),\\
    F_{2}&=2I_{3,f}Q_{f}m_{A^{\prime}}^{2}\\
    &\times[c_{w}^{2}(m_{A^{\prime}}^{2}-m_{Z}^{2})-s_{w}^{2}m_{A^{\prime}}^{2}](m_{A^{\prime}}^{2}+2m_{f}^{2}),\nonumber\\
    F_{3}&=2Q_{f}^{2}[c_{w}^{2}(m_{A^{\prime}}^{2}-m_{Z}^{2})-s_{w}^{2}m_{A^{\prime}}^{2}]^{2}(m_{A^{\prime}}^{2}+2m_{f}^{2}),
\end{align}
which depend on the fermion weak isospin, with ${ I_{3,f}=+1/2}$ for neutrinos and up-type quarks, and ${ I_{3,f}=-1/2 }$ for charged leptons and down-type quarks. The mass of the Z boson is given by ${m_Z=91.1876}$~GeV and its decay width is $\Gamma_{Z}=2.495$~GeV, while ${ c_{w}\equiv \cos\theta_{w} }$ and ${ s_{w}\equiv \sin\theta_{w} }$ are set by the Weinberg angle $\theta_w$.

In the case of low-mass dark photons, the majority of these quarks subsequently form pions and kaons. By combining the individual contributions from these different quark channels, we can focus on the hadronic decay width, which is given by:

\begin{equation}
    \Gamma(A^{\prime}\to hh)=\frac{\alpha \epsilon^{2}}{12}m_{A^{\prime}}|F_{h}(q^{2})|^{2}\Big{(}1-\frac{4m_{h}^{2}}{m_{A^{\prime}}}\Big{)}^{3/2},
    \label{eq:Gamma_Apipi}
\end{equation}

\noindent where $h\equiv \pi, K$, and the form factor $F_{h}$ is taken from Ref.~\cite{Bruch:2004py}, where $q^{2}=m_{A^{\prime}}^{2}$ and $q$ is the total momentum. We note that a massive dark photon cannot decay into 2 photons, even through loop-induced fermion couplings, since it is forbidden by the Landau-Yang theorem~\cite{Landau:1948kw,Yang:1950rg}.

On the other hand, dark photons can decay into a pair of massive $W$-bosons through the $AW^{+}W^{-}$ coupling~\cite{Denner:1991kt}. After summing all the possible polarizations of the $W$-boson final state, the decay rate is:
\begin{align}
    \Gamma(A^{\prime}\to W^{+}W^{-})&=\frac{\alpha \epsilon^{2}}{12}m_{A^{\prime}}\frac{m_{Z}^{4}}{(m_{A}^{2}-m_{Z}^{2})^{2}}\nonumber\\
    \times \Big{(}4\eta_{W}^{4}+16\eta_{W}^{2}& - 17 -3\eta_{W}^{-2}\Big{)}\sqrt{1 - \frac{4m_{W}^{2}}{m_{A^{\prime}}^{2}}},
    \label{eq:Gamma_AWW}
\end{align}
where we have $\eta_{W}=m_{A^{\prime}}/(2m_{W})>1$ for the decay to be kinematically allowed.

Combining all the decay rates in Figure~\ref{fig:DPdecay}, we calculate the branching ratio for each Standard Model decay channel, obtaining results similar to Refs.~\cite{Cirelli:2016rnw, Bai:2016vca, Ekstedt:2016wyi}. The results are shown in Figure~\ref{fig:BRDP}. The $e^{+}e^{-}$-channel is only dominant below the pion mass, while up-type quarks dominate above 300~GeV. All charged lepton branching ratios go to the same values at the high mass limit, and the same is true for all up-type quarks and down-type quarks. This is due to the fact that these fermionic couplings are primarily proportional to the SM QED charges. The enhancement for the quark channels stems from the $N_{c}=3$ colors. We stress that this result is not only applicable to dark photon models, but also any model with vector mediators or dark matter particles that couple to SM fermions in proportion to their electric charges.

\begin{figure}[t]
\centering
\includegraphics[width=0.8\columnwidth]{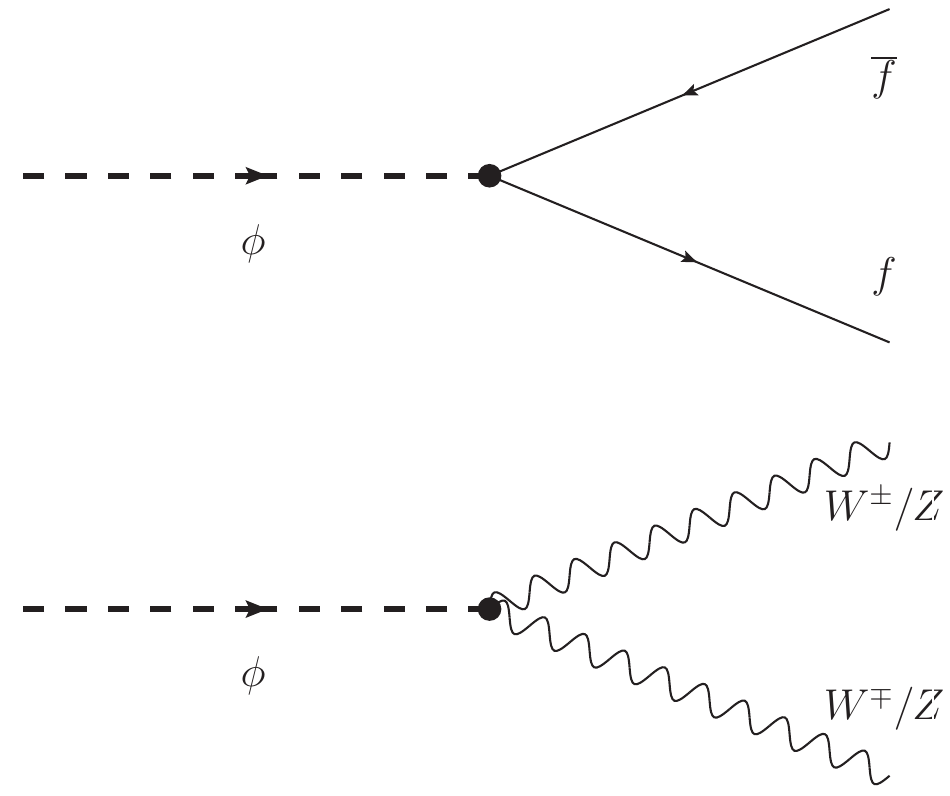}
\caption{Feynman diagrams for scalar dark matter decay to SM particles through Yukawa-like couplings.}
\label{fig:Scalar_decay}
\end{figure}

\subsection{Scalar dark matter}
\label{subsec:scalar}

To construct a scalar dark matter particle, $\phi$, that interacts with other SM particles, we include an interaction with the SM Higgs as $C_{\phi}\phi |H|^{2}$, where $C_{\phi}$ is the coupling strength. When the Higgs gets a vacuum expectation value (VEV), it induces mixing between $\phi$ and the SM Higgs. This implies that the mass eigenstate of $\phi$ strongly depends on the Higgs VEV, and as a result has a scaled down Yukawa coupling. For small enough values of $C_{\phi}$, these couplings can be significantly smaller than the traditional Yukawa couplings, implying that $\phi$ negligibly affects the masses of SM particles.

Thus, similar to the Higgs boson~\cite{ATLAS:2012yve, CMS:2012qbp, Glashow:1961tr, Glashow:1970gm, Weinberg:1967tq}, we can write down the interaction Lagrangian of scalar dark matter with other SM particles after symmetry breaking as~\cite{Pospelov:2008jk}:
\begin{equation}
    \mathcal{L}\supset -\sin \theta \frac{m_{f}}{v}\phi\bar{f}f+ \sin \theta \frac{2 \delta_{V} m_{V}^{2}}{v}\phi V_{\mu}V^{\mu},
    \label{eq:LS}
\end{equation}
where the decay of $\phi$ can proceed to either fermion pairs ($f$) or vector bosons $V\equiv W,Z$.
The SM Higgs VEV is $v\approx246$~GeV, and $\theta$ is the mixing angle between the new scalar $\phi$ and the SM Higgs. Therefore, similar to Higgs couplings~\cite{ParticleDataGroup:2024cfk}, the coefficients $\delta_{W}=1$ and $\delta_{Z}=1/2$.

\begin{figure}[t]
\centering
\includegraphics[width=1\columnwidth]{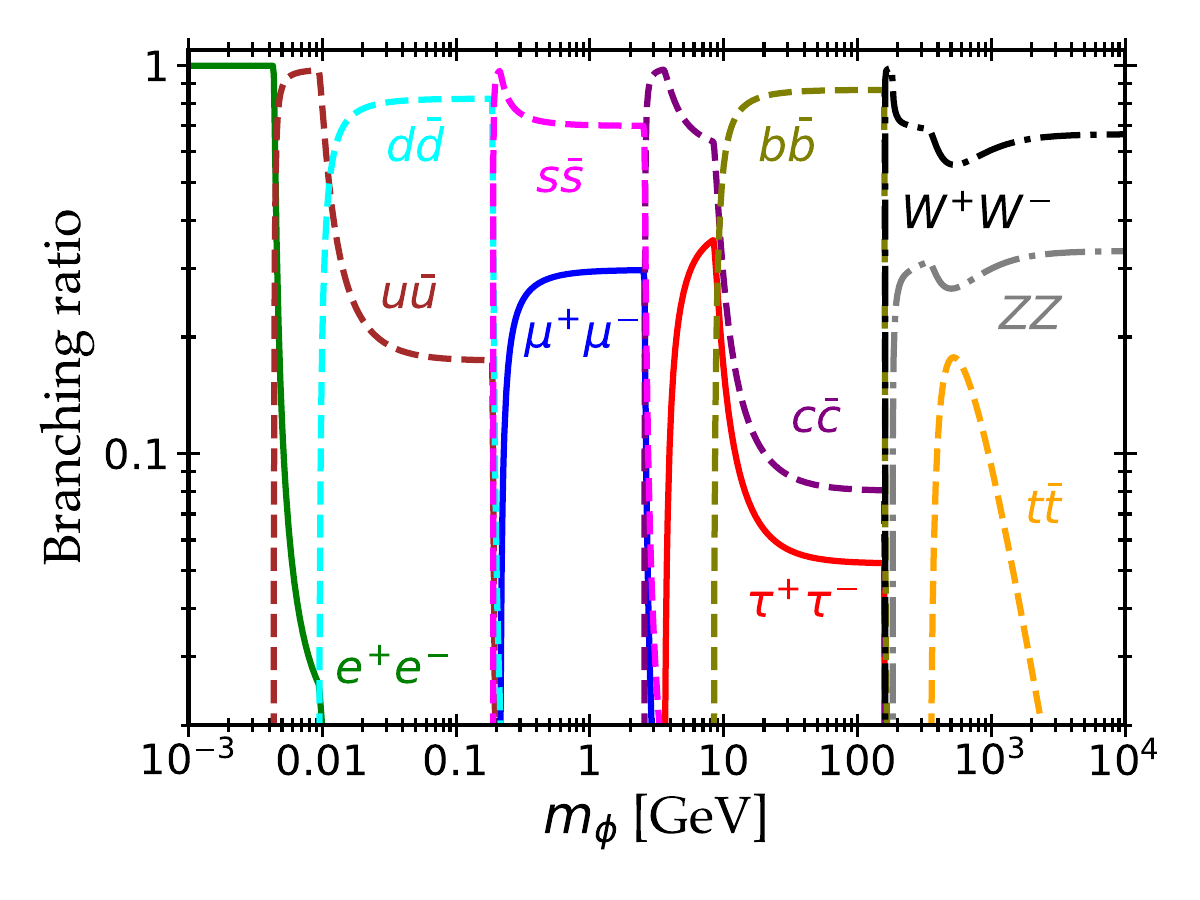}
\caption{Branching ratio of all Standard Model channels for scalar dark matter decay. The solid lines are for charged leptons, while the dashed lines are for quarks. The dash-dotted lines are for $W$ and $Z$ boson production. These results are also applied to scalar boson decays with Yukawa-like couplings.}
\label{fig:BRSc}
\end{figure}

For $m_{\phi}> 1$~MeV, the decay width of scalar dark matter to fermion pairs is given by:
\begin{equation}
    \Gamma(\phi\to \bar{f}f)=\frac{N_{c}G_{F}m_{f}^{2}}{4\sqrt{2}\pi }m_{\phi}\sin^{2}\theta\Big{(}1-\frac{4m_{f}^{2}}{m_{\phi}^{2}}\Big{)}^{3/2},
    \label{eq:Gamma_Sff}
\end{equation}
\noindent where the Fermi constant $G_{F}=1.66\times10^{-5}$~GeV$^{-2}$. Such processes are shown in Figure~\ref{fig:Scalar_decay} (top). On the other hand, the decay width to the $W^{+}W^{-}$ final state is:
\begin{align}
    \Gamma(\phi\to W^{+}W^{-})&=\frac{G_{F}}{8\sqrt{2}\pi}m_{\phi}^{3}\sin^{2}\theta\sqrt{1-\frac{4m_{W}^{2}}{m_{\phi}^{2}}}\nonumber\\
    &\times\Big{(}1-4\frac{m_{W}^{2}}{m_{\phi}^{2}}+12\frac{m_{W}^{4}}{m_{\phi}^{4}}\Big{)},
    \label{eq:Gamma_SWW}
\end{align}
while the $Z$-boson pair final state differs slightly and is given by:
\begin{align}
    \Gamma(\phi\to ZZ)&=\frac{G_{F}}{16\sqrt{2}\pi}m_{\phi}^{3}\sin^{2}\theta\sqrt{1-\frac{4m_{Z}^{2}}{m_{\phi}^{2}}}\nonumber\\
    &\times\Big{(}1-4\frac{m_{Z}^{2}}{m_{\phi}^{2}}+12\frac{m_{Z}^{4}}{m_{\phi}^{4}}\Big{)}.
    \label{eq:Gamma_SZZ}
\end{align}
Both $W^+W^-$ and $Z$-pair decays require a heavy scalar dark matter mass that exceeds twice the vector boson mass. Such diagrams are shown in Figure~\ref{fig:Scalar_decay} (bottom).

Unlike the dark photon case, scalar dark matter opens a new $Z-$boson channel. This neutral boson channel is a distinguishing feature of Yukawa-like-couplings compared to QED-like coupling models. Note that we neglect the di-photon production through fermionic loop-induced diagrams as in the Higgs boson case, due to the suppression of Yukawa couplings in triangle-loop diagrams.

Several important pieces of information can be extracted from Figure~\ref{fig:BRSc}. For dark matter masses between the pion mass up to 161~GeV, quark-final states are dominant for this type of model. This indicates that indirect detection constraints that assume $b\overline{b}$ final states are fairly accurate. However, above twice the weak boson mass, the vector bosons final states are dominant, compared to the $t\overline{t}$ final state, implying that previous work assuming quark final states may provide erroneous results.

\section{Observation data and methodology}
\label{sec:data}

In this section, we calculate the photon and positron fluxes that are produced by dark matter decay, and specifically discuss the observational methodology for calculating x-ray signals observed by INTEGRAL in Subsection~\ref{subsec:integral}, as well as the high energy positron signals observed by AMS-02 in Subsection~\ref{subsec:ams}. 

Throughout this paper, we employ the standard NFW profile~\cite{Navarro:1995iw} in order to calculate the dark matter density throughout the Galactic halo,

\begin{equation}\label{eq: DM profile}
\rho_\text{NFW}(r) = \rho_\text{scale}\left(\frac{r}{R_\text{scale}}\right)^{-\gamma} \left(1 + \frac{r}{R_\text{scale}}\right)^{\gamma-3},
\end{equation}

\noindent where we set the profile index to $\gamma = 1$, the scale radius to $R_\text{scale} = 20$~kpc. We adopt a solar position $R_\odot = 8.5$~kpc and set the scale density $\rho_\text{scale}$ such that the dark matter density at the solar position is $\rho_\odot = 0.4$~GeV/cm$^3$~\cite{Read:2014qva, Salucci:2010qr}.

The photon and positron spectra depend on the dark matter mass and the SM final states. Because we evaluate dark matter decays over a wide range of dark matter masses, there is no single computational package that accurately calculates the resulting photon and positron spectra. Thus, we employ several packages, which provide the most robust results in different mass ranges.

\begin{itemize}
    \item For dark matter masses below 1.5~GeV, we evaluate only the x-ray fluxes from dark matter, using the \texttt{Hazma} package~\cite{Coogan:2019qpu, Coogan:2021sjs, Coogan:2022cdd, Plehn:2019jeo} to calculate x-ray spectra.
    
    \item For dark matter masses from 1.5--10~GeV, there is no existing code that claims to robustly predict the low-energy x-ray flux. To address this gap, we conservatively rely only on the final-state radiation (FSR) component, as this formula remains valid across the entire dark matter mass range. We neglect the additional x-ray flux from dark matter decay, but note that this is subdominant. Consequently, our constraints are conservative, and may potentially improve as new software packages become available that account for decays, which are not currently included in \texttt{Pythia}~\cite{Bierlich:2022pfr}.
    
    \item From 10~GeV to 2~TeV, we find that AMS-02 positron constraints significantly outperform x-ray constraints. We use \texttt{CosmiXs}~\cite{Arina:2023eic}, which accounts for both Electroweak and QCD corrections, to calculate the positron flux. However, we note that both \texttt{PPPC4DMID}~\cite{Cirelli:2010xx, Ciafaloni:2010ti, Baratella:2013fya, Boudaud:2014qra, Buch:2015iya}, and \texttt{DarkSUSY}~\cite{Bringmann:2018lay, Gondolo:2004sc} also provide reasonable results in this mass range.
    
\end{itemize}

Using this combination of spectral and morphological information, we can calculate the relevant dark matter induced fluxes in each experiment.

\subsection{INTEGRAL/SPI: x-ray telescope}
\label{subsec:integral}

The hard x-ray INTErnational Gamma-Ray Astro Physics Laboratory (INTEGRAL) was launched by the European Space Agency (ESA) in~2002. The SPectrometer of INTEGRAL, or SPI, detects photons between 30~keV and 8~MeV. This wide range provides an opportunity to study a multitude of astrophysical phenomena~\cite{Siegert:2015ila, Siegert:2015knp, Siegert:2016ymf, Siegert:2016ijv, Siegert:2019clp, Siegert:2021trw}. Importantly, INTEGRAL provides stringent constraints on dark matter models such as: Axion-Like-Particles (ALP)~\cite{Calore:2022pks, Langhoff:2022bij}, sterile neutrinos~\cite{Calore:2022pks, Langhoff:2022bij, Yuksel:2007xh, Boyarsky:2007ge}, primordial black holes (PBHs)~\cite{Berteaud:2022tws, Iguaz:2021irx, Siegert:2021upf}, and generic WIMP-like decay and annihilation~\cite{Cirelli:2023tnx, Cirelli:2020bpc, DelaTorreLuque:2023huu, DelaTorreLuque:2024zsr}. In the case of dark photon dark matter models, previous studies in Refs.~\cite{Linden:2024fby, Redondo:2008ec, Arias:2012az, Krnjaic:2022wor} only focus on the  mass regime below 1~MeV, where the 3$\gamma$-final state,  called ``the dark photon trident" in Ref.~\cite{Linden:2024fby}, is dominant. At masses that exceed twice the electron mass, 2-body fermion and weak boson final states become kinematically accessible and dominate the decay width, as shown in Figures~\ref{fig:BRDP} and~\ref{fig:BRSc}. This parameter space is not yet explored and is one of the main motivations for our study.

\subsubsection{Astrophysical Background Model}
\label{ssubsect:INTEGRAL_BKG}

\begin{figure}[t]
\centering
\includegraphics[width=\linewidth]{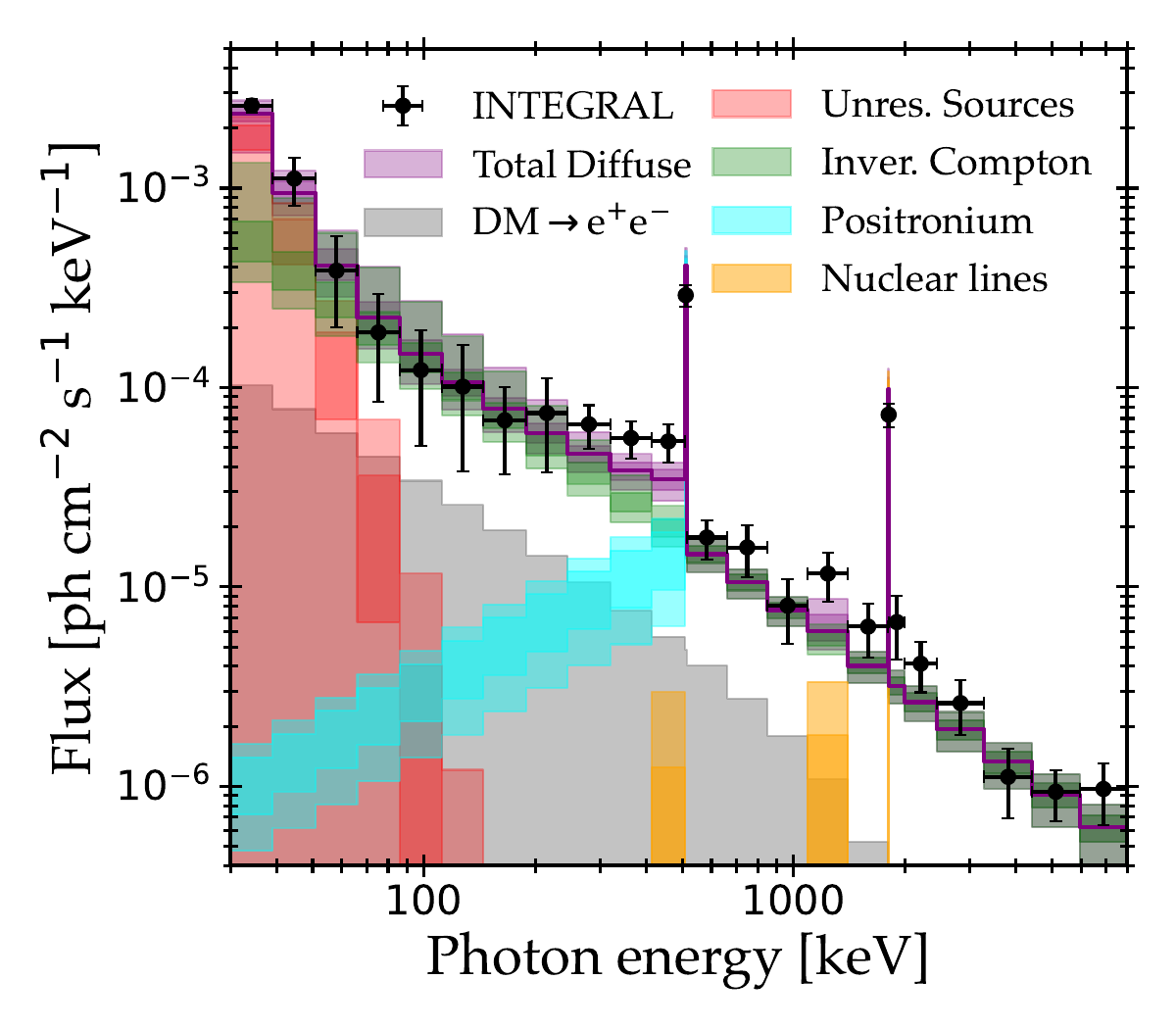}
\caption{INTEGRAL/SPI data compared to a model for the x-ray flux. The 16-year data (black points) spans 30~keV to 8~MeV and is divided into 24~bins. The astrophysical background components include: Unresolved Sources (red), Inverse Compton Scattering (green), Positronium decay (cyan), and Nuclear Lines (orange). Each astrophysical spectrum includes 1$\sigma$ and 2$\sigma$ error bands, and their combined astrophysical diffuse flux is shown in purple. We plot the photon flux from Final State Radiation for a DM decay to electron-positron pairs (gray) at a benchmark bosonic DM mass of $m_{A^{\prime}/\phi}=4$~MeV, (with an arbitrary normalization) with the lifetime $\tau_{\rm DM}=10^{25}$~s.}
\label{fig:INTEGRAL}
\end{figure}

We note that the astrophysical sources are expected to produce the majority of the X-ray flux in this energy range. By including known astrophysical sources, we can produce a much stronger constraint on the dark matter contribution. In Appendix~\ref{appen:0Bkg} we show conservative results where we assume that dark matter produces the entirety of the X-ray data. Based on the framework in Refs.~\cite{Linden:2024fby, Calore:2022pks, Siegert:2021upf}, we use 16-years of INTEGRAL data to investigate dark matter decays. The observed photon flux from 30~keV to 8~MeV is divided into 24~energy bins as shown in Figure~\ref{fig:INTEGRAL}. We note that some bins are narrower than others, in order to describe emission relating to the 511~keV and 1809~keV lines. Using the following background model, we can investigate sub-dominant dark matter contributions. The model contains several astrophysical sources that are shown in Figure~\ref{fig:INTEGRAL}:
\begin{itemize}
    \item Unresolved sources with a power-law plus exponential cutoff spectrum (\textcolor{red}{red}), most likely associated with cataclysmic variables and stars with hot coronae. This emission dominates below 100~keV~\cite{unresolve}.
    \item A power-law spectrum in \textcolor{Green}{green} for the inverse-Compton scattering between GeV cosmic-ray electrons and interstellar radiation fields~\cite{Wang_2020}.
    \item Normalized Positronium decay emission at an energy of 511~keV (\textcolor{cyan}{cyan}) that stems from the annihilation of cool cosmic-ray positrons~\cite{John:2024don} with electrons in the interstellar medium~\cite{Siegert:2015knp}.
    \item Several nuclear lines ($^{7}$Be, $^{22}$Na, $^{60}$Fe) produced by elements that are found in massive stars~\cite{Siegert:2021wlq, Wang_2020}, including, most prominantly, $^{26}$Al, which produces the 1809~keV line~\cite{Diehl:2006cf}. We note that Gaussian distribution in these lines represent uncertainties in the INTEGRAL energy reconstruction.
\end{itemize}

Adding these contributions together, we plot the total astrophysical x-ray spectrum (without any dark matter contribution) in \textcolor{violet}{violet}. All of the distributions, parameters, and best fit values are described in Refs.~\cite{Linden:2024fby, Berteaud:2022tws}.

\subsubsection{Computing Dark Matter Constraints}
\label{ssubsect:INTEGRAL_DM}

Using INTEGRAL/SPI data and this background model, we investigate the dark matter decay scenario for dark matter masses between 1~MeV and 10~GeV. In this mass range, keV--MeV photons from dark matter decay are generated by several scenarios:
\begin{itemize}
    \item Charged lepton final states such as $e^{\pm}$, $\mu^{\pm}$, and charged pions ($\pi^{\pm}$) radiate photons. This process is called Final State Radiation~\cite{Beacom:2004pe, Bystritskiy:2005ib}.
    \item Heavy charged leptons like $\mu^{\pm}$ and $\tau^{\pm}$ decay to lighter states.
    \item Quark final states quickly hadronize and form bound states, that subsequently decay to lighter states, emitting photons.
    \item In models with scalar dark matter, neutral pions ($\pi^{0}$) can be immediately produced and can then decay directly to two-photon final states with a box-like spectrum. The direct production of $\pi^{0}$'s  is forbidden for the spin-1 dark photon dark matter scenarios.
\end{itemize}
All of these channels, whose expressions are provided in Subsection~\ref{appen:photon}, are calculated using the \texttt{Hazma} package for dark matter masses below 1.5~GeV. 

We calculate the x-ray flux that is observed by INTEGRAL as:

\begin{equation}
    \frac{{\rm d}\Phi_{\gamma}}{{\rm d}E_{\gamma}}=\sum_{\rm SM}\frac{\Gamma_{\rm DM \to SM}}{4\pi m_{\rm DM}}\times \frac{{\rm d}N_{{\rm SM}\to \gamma}}{{\rm d}E_{\gamma}}\times D,
    \label{eq:flux}
\end{equation}
\noindent where the $D$-factor, which represents the astrophysical dependence of the photon flux, is given by:

\begin{equation}\label{eq: D factor}
    D=\int {\rm d}\Omega\int_{l.o.s.}{\rm d}s \rho_{\rm NFW}\Big{(}r(s, l, b)\Big{)},
\end{equation}
which is calculated by integrating the dark matter density profile over the line-of-sight (l.o.s.) within the INTEGRAL/SPI region of interest (ROI) defined by longitude $|l|\leq 47.5^{\circ}$ and latitude $|b|\leq 47.5^{\circ}$. We compute the $D-$factor in Eq.~\ref{eq:flux} for dark matter decay as $D=1.29\times 10^{23}$~GeV/cm$^{2}$.

Throughout this mass range, the FSR contribution dominates the dark matter signal in INTEGRAL. This occurs for two different reasons, depending on the dark matter mass. At low masses (below the $\pi^0$ mass), dark matter only decays to $e^+e^-$ pairs, which produce most of their x-ray emission through FSR. In Figure~\ref{fig:INTEGRAL}, we present the arbitrarily normalized x-ray spectrum resulting from the decay of dark matter with a mass of 4~MeV. At higher masses, decays into mesons become dominant. However, these mesons decay into photons with a spectrum that peaks at around half the dark matter mass, exceeding the 8~MeV energy-range of INTEGRAL. The low-energy x-ray flux in these scenarios is still dominated by FSR. We note that the INTEGRAL flux may include additional terms from inverse-Compton scattering and the radiative decay of charged particles produced in the dark matter annihilation event. However, modeling these induces additional systematic concerns as the propagation of these particles must be carefully tracked. Thus, we conservatively choose to model only the FSR contribution from dark matter annihilation. 

At each dark matter mass, we compute the 95\%~CL upper limits on the dark matter decay rate. Specifically, we fit the indirect signal from dark matter decay along with other astrophysical contributions from our background model. Using the \texttt{emcee} MCMC method~\cite{Foreman-Mackey:2012any} that is imported in the \texttt{3ML} package~\cite{Vianello:2015wwa}, we sample all the free parameters for these spectra. The dark matter spectrum is computed as the sum of the spectra from each SM final state multiplied by its branching ratio, which depends on the dark matter models in Figures~\ref{fig:BRDP}~and~\ref{fig:BRSc}. Our analysis is based on the template that is published in Ref.~\cite{Calore:2022pks}, and available publicly via a \href{https://zenodo.org/records/7984451}{Zenodo repository}\footnote{https://zenodo.org/records/7984451}.

\subsection{AMS-02: Cosmic-Ray Positrons}
\label{subsec:ams}

The AMS-02 experiment, located onboard the International Space Station, has provided precise observations of local cosmic-ray fluxes~\cite{Kounine:2012ega}. Due to its great ability to distinguish the charge of incoming cosmic rays, it measures the local cosmic-ray positron flux to great statistical precision over an energy range spanning from a few hundred MeV to about 1~TeV~\cite{AMS:2019rhg}. The positron data has long been used to search for spectral features expected from the annihilation or decay of dark matter particles~\cite{Turner:1989kg, Cirelli:2008jk, Bergstrom:2013jra, John:2021ugy}. Here, we use AMS-02 data to strongly constrain dark photon and scalar dark matter.

\subsubsection{Astrophysical Background Model}\label{sec: astrophysical background model}

We first compute an astrophysical background model of the cosmic-ray positron flux measured at Earth. Unlike photons that travel in straight lines, cosmic-ray positrons are dispersed during propagation, diffusing on magnetic fields due to their electric charge. Accounting for this diffusion requires a more detailed model of the dark matter induced positron flux. Here, we follow Ref.~\cite{John:2021ugy} and use numerical simulations of Galactic cosmic-ray propagation to account for the many processes that affect the positrons during their life.

Specifically, we use the cosmic-ray propagation code \texttt{Galprop v.56}~\cite{Strong:1998pw, Strong:1999sv, Strong:2001fu, Strong:2009xj} to thoroughly model the local positron flux. \texttt{Galprop} simulates cosmic-ray propagation throughout the Galaxy, including a detailed treatment of processes such as cosmic-ray injection and acceleration, diffusion, energy losses and fragmentation. We use the astrophysical background model of Ref.~\cite{John:2021ugy} that provides a precise fit to the local AMS-02 positron data~\cite{AMS:2019rhg}. We provide a brief description of the model here, but refer the reader to Ref.~\cite{John:2021ugy} for more detail.

\begin{figure*}[tbp]
\centering
\includegraphics[width=0.48\linewidth]{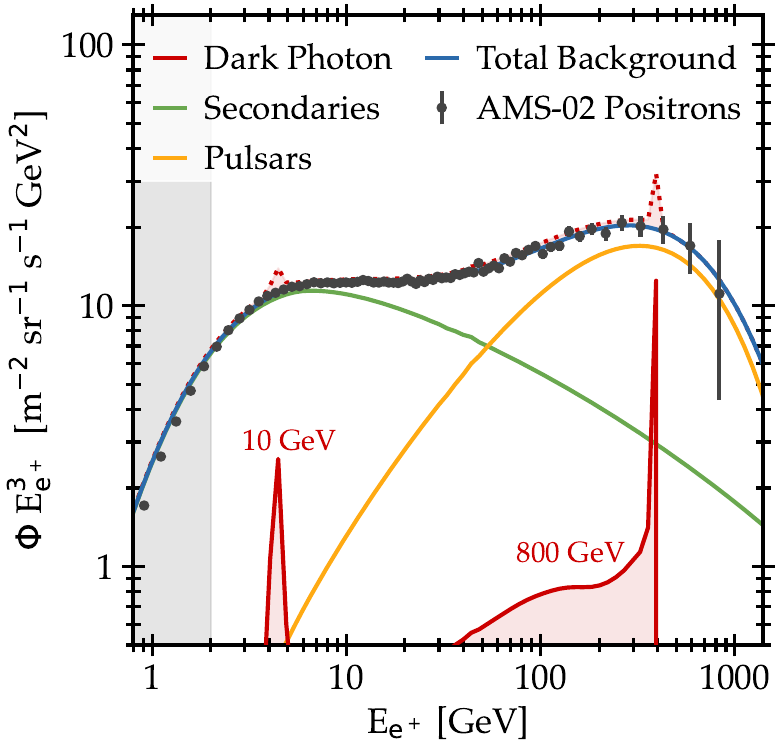}
\hfill
\includegraphics[width=0.48\linewidth]{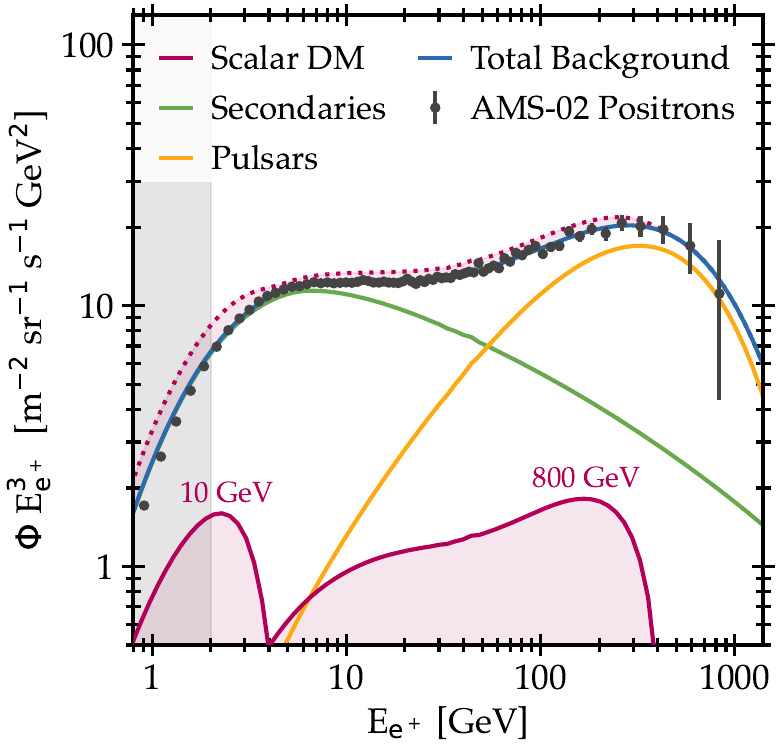}
\caption{The measured AMS-02 positron flux (black points)~\cite{AMS:2019rhg} and its different components. The astrophysical background model, adapted from Ref.~\cite{John:2021ugy} (blue), includes positrons from secondary production (green) and the contribution from a pulsar population that dominates at high energies (orange). Example spectra (at arbitrary flux normalization) from dark matter decays are also shown, including (left panel) the dark photon spectra (red) for dark photon masses of 10~GeV and 800~GeV, and (right panel) the scalar dark matter spectra (magenta) also for masses of 10~GeV and 800~GeV. Dotted lines show the dark matter contribution on top of the background. The light gray shaded region at low energies is not included in the fit due to significant solar modulation uncertainties. In these examples, the spectral features introduced by dark matter decay significantly worsen the fit of the total model to the positron data, and therefore strongly constrain each dark matter model. This is especially clear in the dark photon case, where the highest energies are dominated by the delta-like injection of electron-positron pairs.}
\label{fig: positron flux with contributions}
\end{figure*}

The local cosmic-ray positron flux contains several components that are created by different mechanisms. At low energies ${ \lesssim 30 }$~GeV, positrons are mostly produced as secondaries, \textit{i.e.} in the inelastic interactions of other cosmic rays (mostly protons and Helium) with the interstellar medium. Therefore, the astrophysical background model is also fit to the complementary AMS-02 data for protons~\cite{AMS:2015tnn, AMS:2018qxq} and Helium~\cite{AMS:2015azc} to ensure a realistic model of the secondary positron component.

At high energies (${ \gtrsim 30 }$~GeV), the positron flux is dominated by a new primary component, a result that is known as the \textit{positron excess}. While contributions from annihilating dark matter have been proposed to explain the positron excess in the past (\textit{e.g.},~\cite{Bergstrom:2008gr, Cholis:2008hb, Cholis:2008qq, Cholis:2008wq}), explanations that involve the emission from a sizable population of young pulsars have provided more convincing fits to the data~\cite{Hooper:2008kg, Profumo:2008ms, Linden:2013mqa, Yuksel:2008rf, Malyshev:2009tw, Blasi:2010de}. Recently, with the discovery of highly energetic $\gamma$-ray emission around middle-aged pulsars~\cite{Abeysekara:2017hyn, HAWC:2017kbo, Linden:2017vvb}, called \textit{TeV halos}, the pulsar explanation of the positron excess has been all but settled~\cite{Hooper:2017gtd}.

However, we do not know how many pulsars significantly contribute to the positron flux. Therefore, Ref.~\cite{John:2021ugy} implements a combined model for the total pulsar flux in \texttt{Galprop}, following a positron injection spectrum from Ref.~\cite{Hooper:2008kg} and a model for the galactic distribution of pulsars from Refs.~\cite{Lorimer:2003qc, Lorimer:2006qs}.

The total positron flux from secondaries and pulsars is simulated with \texttt{Galprop} to create the background model. In order to precisely fit AMS-02 data~\cite{AMS:2019rhg}, a large number of free parameters is included in the fit. This includes parameters to describe diffusion, the injection spectra of protons and Helium, the re-acceleration and convection of cosmic rays, as well as the pulsar model. Additionally, solar modulation, \textit{i.e.} the decrease in the cosmic-ray flux below $\sim 10$~GeV due to the Sun's heliosphere, is taken into account by applying the time-, charge- and rigidity-dependent model of Ref.~\cite{Cholis:2020tpi}.

Ref.~\cite{John:2021ugy} computes the astrophysical background model for two different Galactic halo heights. However, as the positrons that contribute to the AMS-02 data, especially at high energies, must have been produced relatively nearby, we do not expect the halo height to have an important impact on our dark matter constraints. This is also pointed out in Ref.~\cite{John:2021ugy}. For our analysis, we only use the default model from Ref.~\cite{John:2021ugy}, which uses a halo height of $z = 5.6$~kpc. This model provides a very accurate fit to the AMS-02 data, with a reduced $\chi^2/{\rm d.o.f.} = 0.88$ for the positron flux.

\subsubsection{Computing Dark Matter Constraints}\label{sec: constraints}
To study dark matter contributions to the positron flux, we add the contributions from dark matter decay into the astrophysical model. We obtain the positron injection spectra for the various decay channels using the \texttt{CosmiXs} code~\cite{Arina:2023eic, Jueid:2023vrb, Jueid:2022qjg, Amoroso:2018qga}, and combine them at each dark matter mass according to Equations~\ref{eq:Gamma_Aff} to \ref{eq:Gamma_AWW} for the dark photon and Equations~\ref{eq:Gamma_Sff} to~\ref{eq:Gamma_SZZ} for scalar dark matter. The result of these equations is shown in the branching ratios in Figures~\ref{fig:BRDP} and~\ref{fig:BRSc}, respectively. We consider bosonic dark matter particles in the mass range of 10~GeV to 2~TeV, and scan dark matter masses over 24 logarithmic mass bins. We add the dark matter contributions into \texttt{Galprop}, including normalizations that are based on the dark matter density profile in Equation~\ref{eq: DM profile}. 

We then re-fit the combined astrophysical and dark matter model to the AMS-02 positron data. We use the fitting procedures provided by \texttt{iMinuit v.1.4.3.}~\cite{iminuit, iminuit_website}. Similar to Ref.~\cite{John:2021ugy}, we take a conservative approach and also include a few parameters in the fit that are particularly important or degenerate with the dark matter flux. Specifically, we vary the diffusion coefficient and diffusion spectral index, as well as 3 parameters that describe the pulsar model (spectral index, spectral cutoff energy and pulsar formation rate), respectively. Then, we compute a $\chi^2$ profile for a range of dark matter fluxes and each dark matter mass. From the $\chi^2$ profile, we calculate the 95\% upper limit (\textit{i.e.} by finding the flux for which ${ \chi^2_\text{95\%} = \chi^2_\text{bgr} + 3.84 }$) to obtain the constraints.

Figure~\ref{fig: positron flux with contributions} shows the local cosmic-ray positron flux with additional contributions from the decay of dark photon dark matter (left panel, red) and scalar dark matter (right panel, magenta). In both cases, the astrophysical model consists of the secondary positrons (green) that are dominant at low energies and a pulsar component (orange) that dominates at high energies. The close match between this model and the AMS-02 positron data (black dots) indicates that astrophysical sources can fit positron observations to percent-level accuracy (see also Ref.~\cite{John:2021ugy}). As an example for the positron contribution from dark matter decay (\textit{i.e.} after propagation), we show spectra for the decay of a 10-GeV and 800-GeV dark matter particle, respectively, for both dark photon and scalar dark matter models. The dotted lines represent the dark matter signals on top of the background model -- which in these examples significantly worsen the fit of the combined background plus dark matter model. Thus, our analysis indicates that AMS-02 positron data produces strong constraints on dark matter decay for both dark photons and scalar dark matter.

We note that dark photon and scalar dark matter have significantly different spectral signatures. This is due to the different branching ratios in each model, as discussed in Section~\ref{sec:model}. In the dark photon case, the positron signal is dominated by a sharp peak near the energy corresponding to half of the dark photon mass. This is produced by the direct decay of the dark photon into electron-positron pairs. Even at hundreds of GeV, when the direct decay to $e^+e^-$ pairs becomes fractionally less important as quark final states become the dominant annihilation pathway, the constraint is still driven by the highly peaked direct injection of positrons. Conversely, the positron spectrum from scalar dark matter is much smoother. This is due to the fact that the coupling to electron-positron pairs is strongly suppressed (see Figure~\ref{fig:BRSc}) based on their low mass. Instead, the signal shape is dominated by decays into heavy quarks, as well as $Z$ and $W$ bosons for scalar dark matter masses above $\sim 161$~GeV.

\begin{figure*}[tbp]
\centering
\includegraphics[width=2\columnwidth]{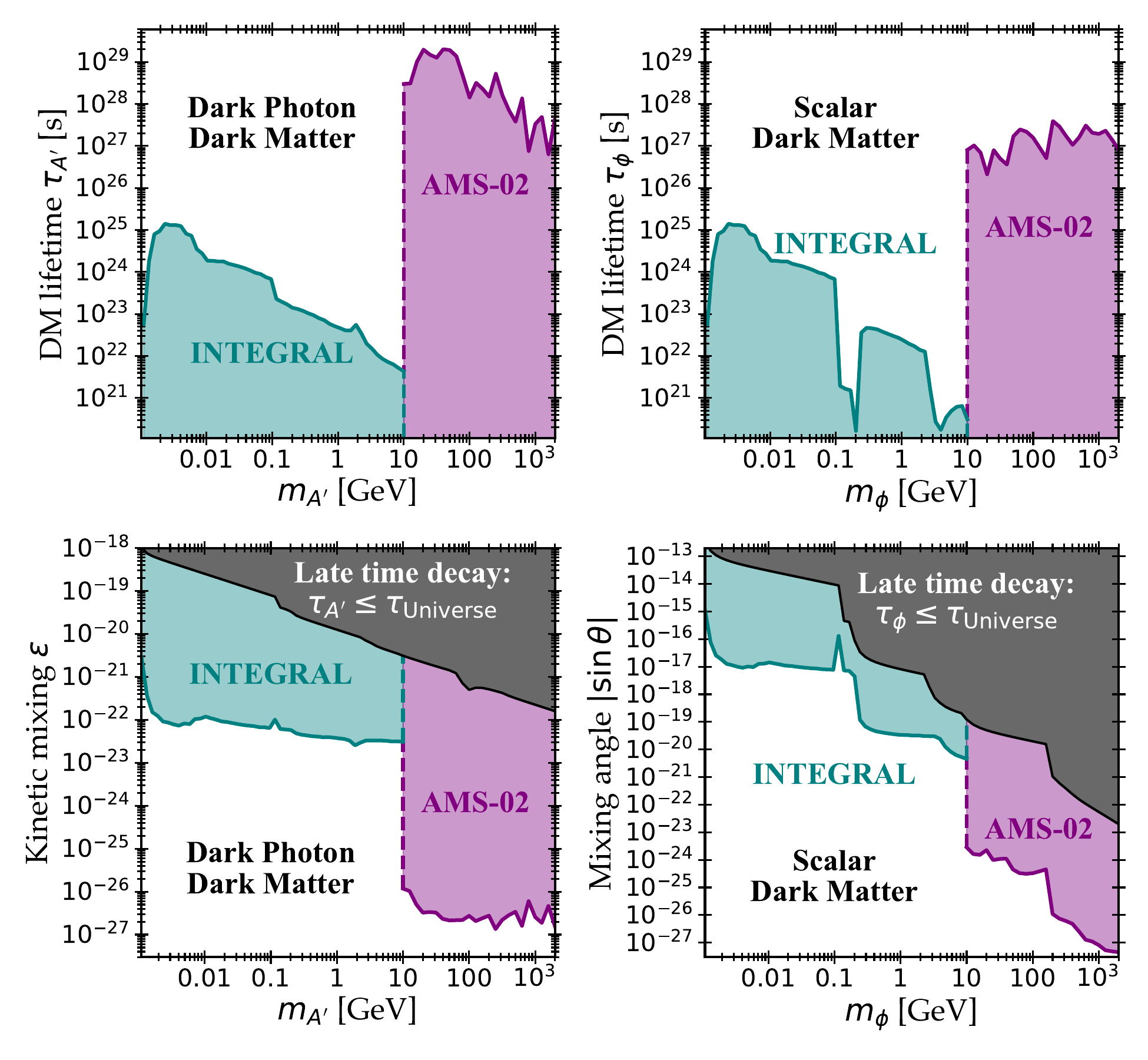}
\caption{INTEGRAL and AMS-02 results on dark matter decay. {\bf Top:} Constraints on the minimum lifetime of dark matter in the mass range from 1~MeV to 2~TeV. The left figure presents results for dark photon dark matter, while the right figure illustrates constraints for scalar dark matter. {\bf Bottom:} New constraints on the dark matter couplings with SM particles in the same mass range. The bottom left figure shows the constraints for dark photon kinetic mixing coupling, while the bottom right figure show the upper limit for mixing angle of the scalar dark matter. The black solid lines and the gray areas demonstrate cosmological constraints for these dark matter models, with $\tau_{U}\approx 4.3\times 10^{17}$~s.}
\label{fig:Final}
\end{figure*}

\section{Constraints on the Dark Matter Lifetime}
\label{sec:result}
Figure~\ref{fig:Final} shows our final constraints on the dark photon and scalar dark matter models, for masses between 1022~keV and 2~TeV. The top panels show the 95\% CL on the minimum decay lifetime, and the bottom panels translate these limits to constraints on the couplings as a function of the dark matter mass. Left panels show the results for dark photon dark matter, and right panels for scalar dark matter. Constraints from INTEGRAL/SPI are shown in \textcolor{teal}{teal} for dark matter masses from 1022~keV to 10~GeV, while results from AMS-02 are shown in \textcolor{violet}{purple} for masses from 10~GeV to 2~TeV. In principle, INTEGRAL can also constrain TeV dark matter, but these results are weaker than AMS-02, due to the fact that the photons observed by INTEGRAL are from FSR at energies below 8~MeV. Conversely, AMS-02 limits stem from observations near the dark matter mass that include most of the final state energy. On the other hand, because of the high uncertainty in the positron flux caused by solar modulation below $\sim$5~GeV, we only consider dark matter masses above 10~GeV for AMS-02 constraints.

For the dark photon, INTEGRAL/SPI reaches constraints from $\sim 10^{22}$~s to $\sim 10^{25}$~s, while AMS-02 pushes the limit up to $\sim 10^{29}$~s near 10~GeV, before falling off for higher masses. Meanwhile for scalar dark matter, the constraints from INTEGRAL span from $10^{21}$--$10^{25}$~s, while the AMS-02 limit fluctuates between $10^{26}$--$10^{27}$~s.

\begin{figure*}[t]
\centering
\includegraphics[width=2\columnwidth]{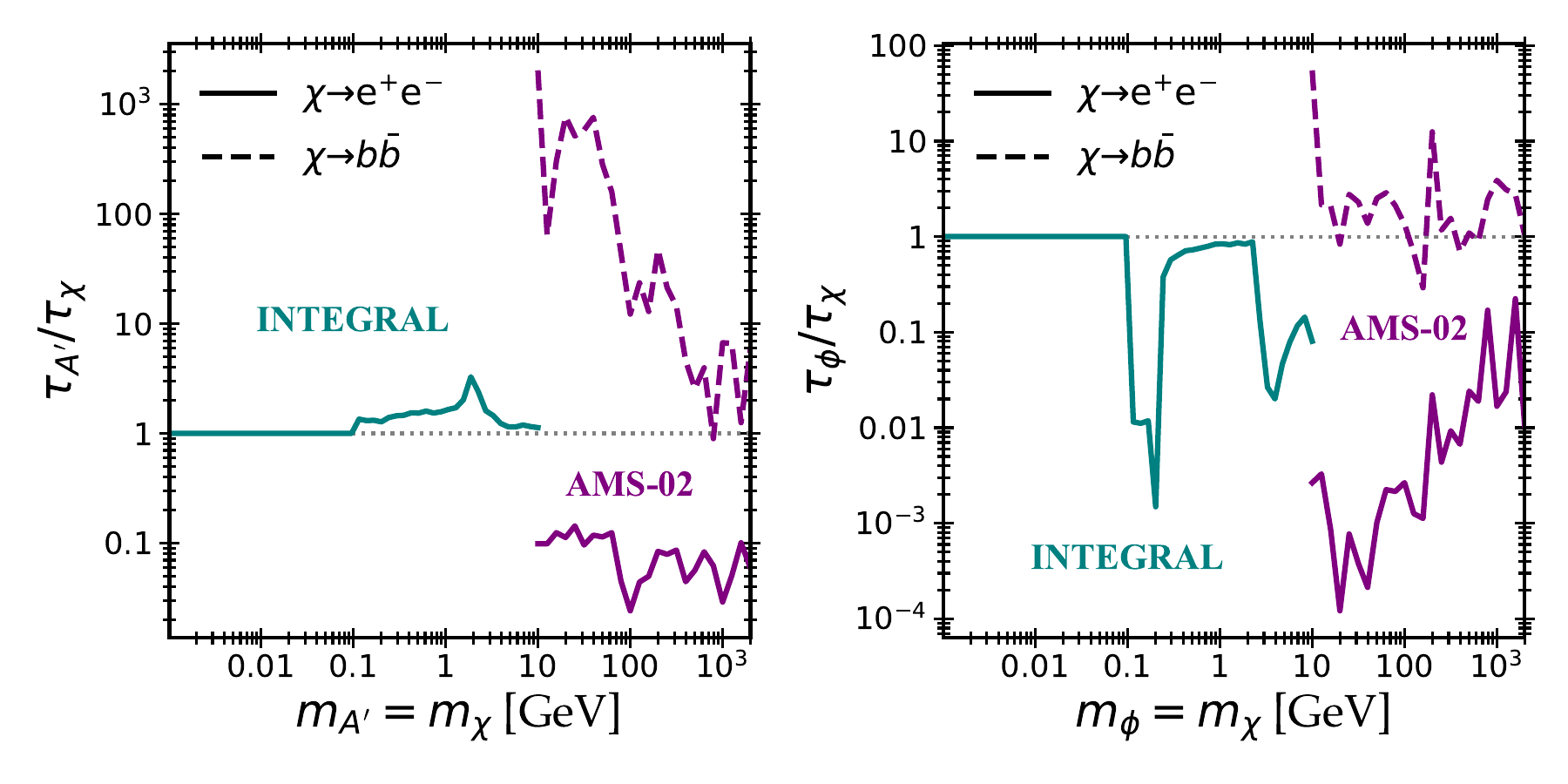}
\caption{The change in the dark matter lifetime constraints when generic dark matter $\chi$ decay final states to $e^+e^-$ (solid lines) or $b\bar{b}$ (dashed lines) are assumed compared to calculating the complete couplings for dark photon dark matter ({\bf left}), and scalar dark matter ({\bf right}). The lifetime ratio using INTEGRAL are in teal, while AMS-02 results are in purple. Assuming simple final states, considering only a single decay channel, produces significant errors in calculating the constraints on the dark matter lifetime, compared to more realistic models.}
\label{fig:ratio}
\end{figure*}

In both cases INTEGRAL limits become weaker above the pion mass, when the dark matter begins to decay into quarks and mesons, which suppresses the final state radiation from charged lepton production. Compared to the dark photon, the INTEGRAL constraints on scalar dark matter is decreased more significantly because the branching ratio of Yukawa-like-couplings to charged leptons are also much smaller compare to the charged (and weak isospin) couplings for the dark photon. 

For AMS-02, dark photon decay produces a sharp spectral feature near half the dark matter mass, as shown in Figure~\ref{fig: positron flux with contributions}. This produces very strong constraints, as the AMS-02 positron data is smooth. We note that this effect is strongest in the range between  ${ \mathcal{O}(10-100) }$~GeV. where the AMS-02 data is most precise. Conversely, for scalar dark matter in this mass range, the positron spectra are much smoother, resulting in weaker constraints compared to the dark photon case. This stems primarily from the fact that the smoother scalar dark matter spectrum is more degenerate with the pulsar contribution, allowing the pulsar model to float and weaken the limits. However, at very high masses, the constraints between dark photon and scalar dark matter become similar, due to the fact that scalar dark matter distributes a larger fraction of its decay power at lower energies, where AMS-02 data is better.

We use these constraints on the dark matter lifetime to calculate upper limits for the relevant dark matter couplings in the bottom panels of Figure~\ref{fig:Final}. We constrain the kinetic mixing of a dark photon (left), and the mixing angle between scalar dark matter with the SM Higgs boson (right). We compare these limits to cosmological constraints that stem from the observation that the dark matter lifetime must exceed the age of the Universe $\tau_{U}\approx 4.3\times 10^{17}$~s.

For dark photon kinetic mixing, $\epsilon$, INTEGRAL constraints exceed ~10$^{-22}$ across the 1~MeV--10~GeV mass range, which is 2--4 orders of magnitude stronger than cosmological constraints. The AMS-02 results further push this coupling limit down to $\sim$$10^{-26}$--$10^{-27}$, which is a 5--7 order of magnitude improvement. We note that the significant gap in the sensitivity between MeV-scale instruments and the GeV-scale constraints offered by AMS-02 motivate additional search techniques~\cite{Cirelli:2023tnx, Lopez-Honorez:2013cua, Slatyer:2015jla, Wadekar:2021qae, Liu:2016cnk, Liu:2020wqz}, and next-generation MeV-scale instruments~\cite{AMEGO:2019gny, ODonnell:2024aaw}.

For scalar dark matter, constraints on the absolute value of the mixing angle $|\theta|\approx |\sin\theta|$ with INTEGRAL span from $10^{-16}$ at 1~MeV up to $10^{-20}$ at 10~GeV, which are stronger than cosmological constraints by 3--4 orders of magnitude.  This bound significantly weakens around the pion mass, due to the suppression of electron production and its subsequent final state radiation. The AMS-02 data strengthens this bound to $10^{-24}$--$10^{-27}$ in the 10~GeV--2~TeV mass range, with a sudden improvement around 161~GeV. This threshold opens the production of $W$ and $Z$ bosons, which subsequently directly decay to $e^+e^-$ with a relatively sharp spectrum. Overall, indirect detection searches improve the upper limit on Yukawa-like mixing angles from between 3--6 orders of magnitude compared to the cosmological constraint.

\begin{figure}[t]
\centering
\includegraphics[width=1\columnwidth]{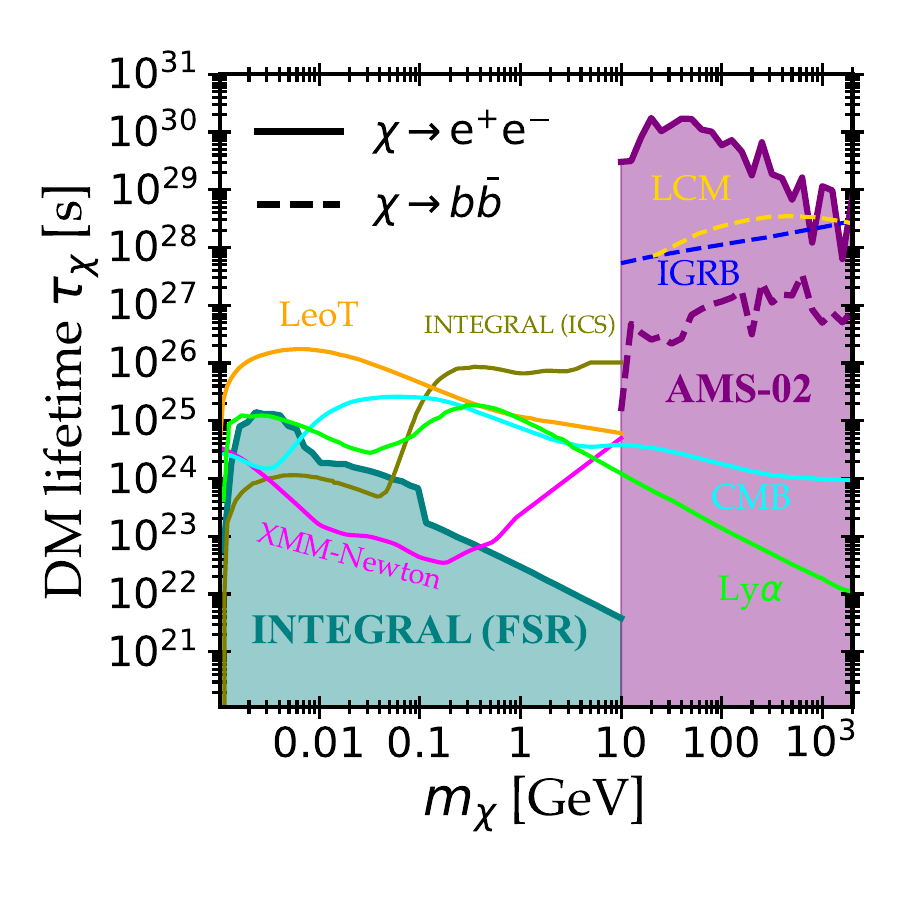}
\caption{INTEGRAL with FSR (teal) and AMS-02 (purple) constraints on dark matter lifetime, with 100\% decay channel: solid lines for e$^{+}$e$^{-}$ final state, and dashed lines for $b\bar{b}$ final state. We compare with previous results on dark matter decays for these two final states: INTEGRAL constraints based on the ICS signal (olive)~\cite{Cirelli:2023tnx}, XMM-Newton (magenta)~\cite{Cirelli:2023tnx, Balaji:2025afr}, LeoT (orange)~\cite{Wadekar:2021qae}, Lyman-$\alpha$ (lime)~\cite{Liu:2020wqz}, CMB (cyan)~\cite{Liu:2016cnk}, Large Magellanic Cloud (gold)~\cite{Regis:2021glv}, Isotropic Gamma-Ray Background (blue)~\cite{Blanco:2018esa}. }
\label{fig:CompareLifetime}
\end{figure}

We also apply our analysis to constrain the dark matter lifetime within the standard (but theoretically unlikely) context of dark matter decay into single channels: $e^{+}e^{-}$ and $b\bar{b}$. Our results are shown in Figure~\ref{fig:CompareLifetime}, with a full discussion in Sec.~\ref{sect:single}. In Figure~\ref{fig:ratio} we show the ratio of the minimum allowed dark matter lifetime for dark photon and scalar dark matter compared to these single decay models. 

Below the pion mass, dark matter can only decay to neutrinos and $e^+e^-$. Thus the constraint for the single $e^{+}e^{-}$-final states is applicable for most models. Above the QCD scale, the FSR of $e^{+}e^{-}$, $\mu^{+}\mu^{-}$, and charged pions produces a bright x-ray flux. Thus, the x-ray constraints on the electric coupling of the dark photon are slightly stronger than for the single decay to $e^+e^-$. For scalar dark matter with Yukawa couplings, the $e^{+}e^{-}$ channel is suppressed, so the constraint is weaker than in the single decay model. 

Above 10~GeV, AMS-02 constraints on both dark photon and scalar dark matter are suppressed compared to the direct production of $e^{+}e^{-}$, by 1 order of magnitude for dark photons, but 3--4 orders of magnitude for scalar dark matter. Comparing our results to decay into $b\bar{b}$, both the dark photon and scalar dark matter constraints are stronger because they produce $e^{+}e^{-}$ directly. Our results demonstrate that realistic decay constraints can vary by orders of magnitude compared to the simple decay channels commonly used in indirect detection.

\section{Single Channel Limits}
\label{sect:single}

While a critical point of this paper is that accurate limits should include the full branching ratios for each dark matter decay process, literature comparisons are currently only possible under the assumption that dark matter decays via a single decay channel. Thus, in this section only, we show the INTEGRAL and AMS-02 constraints calculated following the same method utilized in previous sections, but analyzing dark matter models that decay only via single decay channels: $e^{+}e^{-}$ (solid) and $b\bar{b}$ (dashed).

The INTEGRAL data constrains 1~MeV to 10~GeV dark matter only for the $e^{+}e^{-}$ FSR channel, while AMS-02 looks for positron signals coming from both channels for dark matter with masses above 10~GeV. We compare our results with previous works that consider these two decay channels~\cite{Cirelli:2023tnx, Wadekar:2021qae, Liu:2020wqz, Liu:2016cnk, Regis:2021glv, Blanco:2018bbf}. We find that our results set world leading constraints for dark matter decay into $e^+e^-$ pairs for masses between 10~GeV and 2~TeV, significantly surpassing previous CMB and Lyman-$\alpha$ constraints. The excess strength of these limits, along with the fact that the b$\bar{b}$ channel is typically kinematically dominant, implies that our constraints would likely also exceed any previous limits which were recomputed with realistic branching fractions.

While our constraints on dark matter decay into $e^+e^-$ pairs below 10~GeV are weaker than limits from other works, we note that there are significant difference in the astrophysical and cosmological assumptions utilized in these results. For example, compared to the CMB and Ly-$\alpha$ constraints, our results test relatively local dark matter at a redshift of $z=0$.

We note that our constraints can be exceeded by results from LeoT, XMM-Newton  and especially the previous ICS analysis from INTEGRAL~\cite{Cirelli:2020bpc}, which is the strongest for DM masses between $\sim$100~MeV and $\sim$10~GeV. However, our constraints are unique in that they are independent of any assumptions regarding cosmic-ray diffusion and astrophysical magnetic fields within the Galaxy -- since we study only the FSR which is immediately produced by dark matter decay, thus make our results are conservative. Several recent works~\cite{Cirelli:2023tnx, DelaTorreLuque:2023olp} found that XMM-Newton constraints stemming from the inverse-Compton scattering of the $e^+e^-$ pairs produced by dark matter decay could exceed these limits. However, more recent work has found that these papers included an error related to the misrepresentation of the exposure time of XMM-Newton observations. More recent studies have corrected this issue~\cite{Balaji:2025afr, Cirelli_2025}. The constraints from LeoT depend on the heating of gas within this dwarf galaxy, and thus sensitively depend on the efficiency of cosmic-ray diffusion and the gas dynamics of LeoT. Our constraints for direct decay into $b\bar{b}$ are weaker than $\gamma$-ray constraints, which can directly probe photon production from bottom quark hadronization.

\section{Summary and outlook}
\label{sec:summary}
In this \textit{paper}, we derive strong constraints on the decay of dark photon and scalar dark matter models, using x-ray data from INTEGRAL/SPI and cosmic-ray positron data from AMS-02. For both models, we consider the mass range from 1~MeV to 2~TeV. For dark matter masses below 10~GeV we derive our limits using x-ray data, while for higher masses we use positron data. We obtain constraints on the decay lifetime from $10^{22}$--$10^{29}$~s for dark photon dark matter, and from $10^{21}$--$10^{27}$~s for scalar dark matter. In general, our constraints for these two common dark matter models are about 4--12 orders of magnitude longer than the age of the Universe.

For both models, we provide a detailed calculation of the branching ratios into all possible SM states. These branching ratios also can be applied to other beyond-Standard Model scenarios, for either vector couplings that depend on quantum numbers such as the electric charge and isospin, or scalar couplings that are proportional to the particle mass. Using the branching ratios for all SM decay modes across the 1~MeV--2~TeV mass range, we derive both cosmological and indirect detection constraints on the relevant couplings of each model. For dark photon dark matter, we set constraints on kinetic mixing spanning from $10^{-21}$--$10^{-22}$ from $\sim 1$~MeV--10~GeV, and from $10^{-26}$--$10^{-27}$ from 10~GeV--2~TeV. For scalar dark matter, we set constraints on the mixing angle between $ 10^{-16}$--$10^{-20}$ from 1~MeV--10~GeV, and from $10^{-24}$--$10^{-27}$ from 10~GeV to 2~TeV.

Although our results only account for two relatively simple beyond-Standard Model particles, our constraints can serve as benchmark points for studies that consider more complex dark matter model building~\cite{Nguyen:2025tkl, Cirelli:2025rky}. Finally, we note that the huge gain in sensitivity obtained by deploying and utilizing current generation GeV-scale instrumentation (like the Fermi-LAT and AMS-02) motivates additional work in similarly utilizing and developing next-generation X-ray and MeV $\gamma$-ray instrumentation (such as COSI~\cite{Tomsick:2021wed}, AMEGO~\cite{AMEGO:2019gny} and e-ASTROGAM~\cite{e-ASTROGAM:2017pxr}), which can more sensitively distinguish the spectrum and morphology of any putative dark matter signals from astrophysical backgrounds~\cite{ODonnell:2024aaw}. It is worth noting that the strength of our results in Figure~\ref{fig:INTEGRAL} depends primarily on the uncertainty in astrophysical background modeling. Thus, observational and theoretical advances that allow us to decrease the uncertainty in the astrophysical background model (most notably from inverse-Compton scattering and unresolved sources), will provide more significant advances to dark matter constraints than simply improving the flux sensitivity of next-generation instrumentation. This motivates the development of instruments with improved angular resolution and spectral coverage.

\section*{Acknowledgement}
\vspace{-0.4cm}
We would like to thank Carlos Blanco, Francesca Calore, Ariane Dekker, Ciaran O'Hare, Phan Anh Vu, and Thomas Siergert for fruitful discussions. We thank Mattia Di Mauro for assistance with  \texttt{CosmiXs}, and Logan Morrison with help setting up \texttt{Hazma}. T.T.Q.N and TL acknowledge support by the Swedish Research Council under contract 2022-04283. T.T.Q.N
is also grateful for the support by the COST Action COSMIC WISPers CA21106, supported by COST (European Cooperation in Science and Technology). IJ acknowledges support from the Research grant TAsP (Theoretical Astroparticle Physics) funded by INFN, and Research grant ``Addressing systematic uncertainties in searches for dark matter'', Grant No.\ 2022F2843L, CUP D53D23002580006 funded by the Italian Ministry of University and Research (\textsc{mur}). The work of T.M.P.T. is supported in part by the US National Science Foundation under Grant PHY-2210283. 

This work made use of {\tt Numpy}~\cite{Harris_2020}, {\tt SciPy}~\cite{Virtanen:2019joe}, {\tt astropy}~\cite{Astropy:2013muo}, {\tt matplotlib}~\cite{HunterMatplotlib}, {\tt Jupyter}~\cite{2016ppap.book...87K}, {\tt Jaxodraw}~\cite{Binosi:2008ig}, \texttt{3ML}~\cite{Vianello:2015wwa}, \texttt{Mathematica}~\cite{Mathematica}, \texttt{PackageX}~\cite{Patel:2015tea}, as well as {\tt Webplotdigitizer}~\cite{Rohatgi2022}.

\appendix

\section{Photon energy spectrum}
\label{appen:photon}

There are two scenarios where photons can be produced by dark matter decay. Stable and long-lived final states such as $e^{\pm}$ and $\mu^{\pm}$ (along with charged pion $\pi^{\pm}$), can radiate photons directly, without changing the nature of the particle. This process is called Final State Radiation. Leptons that are heavier than the electron $\mu^{\pm}$, $\tau^{\pm}$, on the other hand, decay to photons along with lighter leptonic states. Additionally, charged pions, mesons and baryons that are hadronized from the direct production of quarks decay to produce photons. In some special cases, such as the neutral pion production, the $\pi^{0}$ can decay directly into two photons. 

\subsection{Final State Radiation}
\label{subappen:fsr}

For $\ell^{\pm}\equiv e^{\pm}, \mu^{\pm}$ final states, the FSR spectrum from dark matter decay is given by:
\begin{equation}
\begin{split}
    \frac{{\rm d}N^{\ell^{\pm}}_{\gamma}}{{\rm d}E_{\gamma}}\Big{|}_{\rm FSR}&=\frac{2\alpha}{\pi \beta(3-\beta^{2})m_{\rm DM}}\\
    &\times\Big{[}\mathcal{A}\times\ln\frac{1 + R(\nu)}{1-R(\nu)}-2\mathcal{B}\times R(\nu)\Big{]},
\end{split}
\end{equation}
where the variables and functions depend on the initial and final masses of the decay as:
\begin{align}
    \nu &= 2E_{\gamma}/m_{\rm DM},\quad \mu  = m_{\ell}/m_{\rm DM},\\
    \beta^{2} & = 1-4\mu^{2}, \quad R(\nu) =\sqrt{1-\frac{4\mu^{2}}{1-\nu}},\\
    \mathcal{A}&=\frac{(1+\beta^{2})(3-\beta^{2})}{\nu}-2(3-\beta^{2})+2\nu,\\
    \mathcal{B}&=\frac{3-\beta^{2}}{\nu}(1-\nu)+\nu,
\end{align}
which are modified from Refs.~\cite{Cirelli:2020bpc, Bystritskiy:2005ib} in order to model dark matter decay, instead of annihilation. Similarly, the FSR spectrum stemming from the production of charged pions is given by:
\begin{align}
        \frac{{\rm d}N^{\pi^{\pm}}_{\gamma}}{{\rm d}E_{\gamma}}\Big{|}_{\rm FSR}=\frac{4\alpha}{\pi \beta m_{\rm DM}}\Big{[}&\Big{(}\frac{\nu}{\beta^{2}}-\frac{1-\nu}{\nu}\Big{)}R(\nu)\\
        &+\Big{(}\frac{1+\beta^{2}}{2\nu}-1\Big{)}\ln\frac{1+R(\nu)}{1-R(\nu)}\Big{]}.\nonumber
\end{align}

\subsection{Radiative decay}
\label{subappen:decay}

In addition to FSR, muons can decay into electrons and directly produce soft photons through the radiative decay channel $\mu^{-}\to e^{-}\bar{\nu}_{e}\nu_{\mu}\gamma$. Following the notation from Refs.~\cite{Coogan:2019qpu, Coogan:2021sjs, Kuno:1999jp, Essig:2009jx}, the decay spectrum in the muon rest frame has the analytical form given by:
\begin{align}
    \frac{{\rm d}N^{\mu}_{\gamma}}{{\rm d}E_{\gamma}}\Big{|}_{\rm Dec}^{E_{\mu}=m_{\mu}}=\frac{\alpha(1-x)}{36\pi E_{\gamma}}\Big{\{}&12[3-2x(1-x)^{2}]\log\Big{(}\frac{1-x}{r}\Big{)}\nonumber\\
    +x(1-&x)(46-55x)-102\Big{\}},
\end{align}
where $x=2E_{\gamma}/m_{\mu}$, $r=(m_{e}/m_{\mu})^{2}$. The maximum photon energy for this radiative decay in the muon rest frame is $E^{\rm max}_{\mu}=m_{\mu}(1-r)/2\simeq 52.8$~MeV.

Similarly, charged pions can also undergo radiative decay, for example, $\pi^{-}\to \ell^{-}\bar{\nu}_{\ell}\gamma$ with $\ell\equiv e,\mu$. The photon spectrum in the pion rest frame is expressed as ~\cite{Bryman:1982et}:
\begin{equation}
    \frac{{\rm d}N^{\pi}_{\gamma}}{{\rm d}E_{\gamma}}\Big{|}^{E_{\pi}=m_{\pi}}_{\rm Dec}=\frac{\alpha[f(x)+g(x)]}{24\pi m_{\pi}f^{2}_{\pi}(r-1)^{2}(x-1)^{2}rx},
\end{equation}
where $x=2E_{\gamma}/m_{\pi}$, $r=(m_{\ell}/m_{\pi})^{2}$, similar to muon decay, and the pion decay constant is $f_{\pi}=92.2$~MeV. The two functions, $f$ and $g$, relate to axial and vector currents and are calculated as:
\begin{align}
    f(x)&=(r+x-1)\\
    \times&\Big{\{}m_{\pi}^{2}x^{4}(F_{A}^{2}+F_{V}^{2})[r^{2}-rx+r-2(x-1)^{2}]\nonumber\\
    &-12\sqrt{2}f_{\pi}m_{\pi}r(x-1)x^{2}[F_{A}(r-2x+1)+xF_{V}]\nonumber\\
    &-24f_{\pi}^{2}r(x-1)[4r(x-1)+(x-2)^{2}]\Big{\}},\nonumber
\end{align}
\begin{align}
    g(x)&=12\sqrt{2}f_{\pi}r(x-1)^{2}\log\Big{(}\frac{r}{1-x}\Big{)}\\
    \times&\Big{\{}m_{\pi}x^{2}[F_{A}(x-2r)-xF_{V}]\nonumber\\
    &+\sqrt{2}f_{\pi}(2r^{2}-2rx-x^{2}+2x-2)\Big{\}},
\end{align}
with the axial and vector form factors given by:
\begin{align}
    &F_{A}=0.119,\\
    &F_{V}(q^{2})=F_{V}(0)(1+aq^{2}),\\
    &F_{V}=0,\quad a=0.1,\quad q^{2}=(1-x).
\end{align}

The total radiative decay of a charged pion in its rest frame is calculated as:
\begin{equation}
    \begin{split}
        \frac{{\rm d}N^{\pi}_{\gamma}}{{\rm d}E_{\gamma}}\Big{|}^{E_{\pi}=m_{\pi}}_{\rm RadTot}=&\sum\limits_{\ell=e,\mu}{\rm BR}(\pi\to \ell \nu_{\ell})\frac{{\rm d}N^{\pi}_{\gamma}}{{\rm d}E_{\gamma}}\Big{|}^{E_{\pi}=m_{\pi}}_{\rm Dec}\\
        &+{\rm BR}(\pi\to \mu \nu_{\mu})\frac{{\rm d}N^{\mu}_{\gamma}}{{\rm d}E_{\gamma}}\Big{|}_{\rm Dec}^{E_{\mu}=E_{\star}},
    \end{split}
\end{equation}
where $E_{\star}=(m_{\pi}^{2}+m_{\mu}^{2})/(2m_{\pi})$, indicates that pion decay can also produce on-shell muons that subsequently decay to low energy photons as discussed above.

For neutral pions, the only decay channel is di-photon production, which produces a sharp line in the pion rest frame as:
\begin{equation}
     \frac{{\rm d}N^{\pi^{0}}_{\gamma}}{{\rm d}E_{\gamma}}\Big{|}^{E_{\pi}=m_{\pi}}_{\rm \pi^{0}\to\gamma\gamma}=2\times\delta(E_{\gamma}-\frac{m_{\pi^{0}}}{2}).
\end{equation}

\begin{figure}[t]
\centering
\includegraphics[width=1\columnwidth]{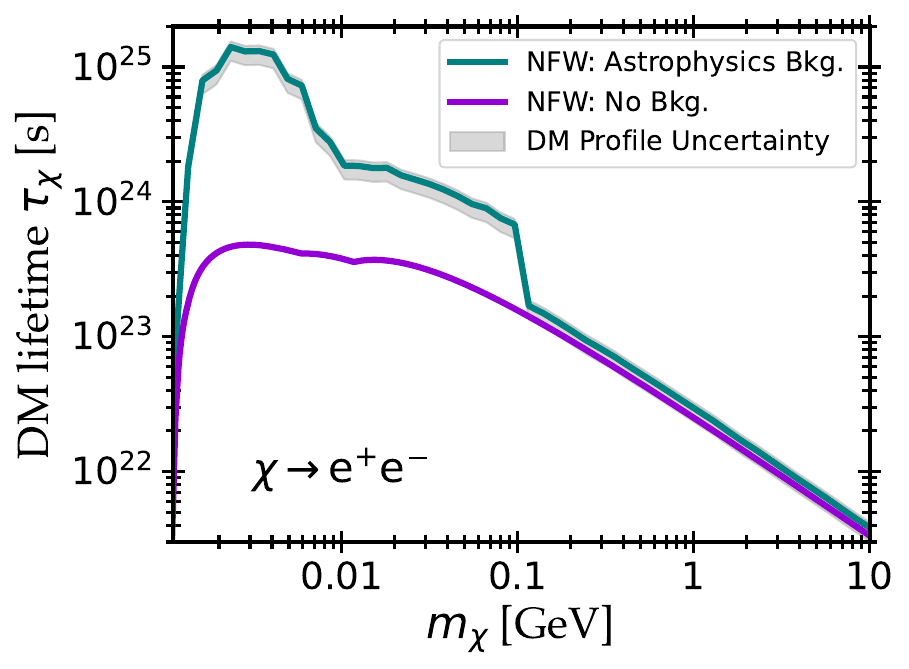}
\caption{INTEGRAL/SPI constraints on the DM decay lifetime for the $\chi\to e^{+}e^{-}$ channel. The teal and purple curves show our limits assuming the NFW profile, with and without modeling the astrophysical background, respectively. The gray bands span the envelope ontained by varying the DM density profile across the NFW, Einasto~\cite{Navarro:2008kc}, Burkert~\cite{Burkert:1995yz}, and Moore~\cite{Diemand:2004wh} choices for the astrophysical background scenario.}
\label{fig:NoBgk}
\end{figure}

To obtain the decay spectrum in the lab-frame, a Lorentz boost is applied to these radiative decay spectra, where $E_{\mu/\pi}=m_{\rm DM}/2$. All spectra and boost transformations are handled within the {\tt Hazma} package, which generates accurate radiative decay spectra for dark matter decays with masses up to 1.5~GeV~\cite{Coogan:2019qpu, Coogan:2021sjs} (for the computation of higher masses, see the main text).

%\section{Dark matter models with single decay channels}
%\label{appen:dmgeneric}

\vspace{0.1cm}

\section{Systematic uncertainties in the X-ray constraints}
\label{appen:0Bkg}

%We derive the INTEGRAL/SPI constraints on the DM decay lifetime for the generic e$^{+}$e$^{-}$ channel, assuming no astrophysical background contribution as described in Sect.~\ref{ssubsect:INTEGRAL_BKG}. This conservative limit is obtained by requiring that the predicted photon flux from DM decay does not exceed the observed photon count in any energy bin of the INTEGRAL observation. In Fig.~\ref{fig:NoBgk}, we show both the constraints with and without background modeling for comparison. Including the astrophysical background model improves the lower bound on the DM decay lifetime by approximately two orders of magnitude for DM masses below 100~MeV, reducing to a factor of a few at higher masses. Since the background model primarily affects the constraint below the QCD scale, where only the e$^{+}$e$^{-}$ channel is available, our X-ray constraints on dark photon and scalar DM particles above 100~MeV remain unaffected.

While astrophysical emission is very likely to dominate the x-ray emission in the INTEGRAL/SPI energy band, one may make a conservative assumption that dark matter produces the entirety of the observed x-ray flux. Based on this, we can derive the INTEGRAL/SPI constraints on the DM decay lifetime for the generic $e^{+}e^{-}$ channel, assuming no astrophysical background contribution as described in Sec.~\ref{ssubsect:INTEGRAL_BKG}. We obtain the conservative limit by requiring that the predicted photon flux from DM decay does not exceed the observed photon count in any energy bin of the INTEGRAL observation.

In Fig~\ref{fig:NoBgk}, we show the resulting constraint based on the x-ray flux for models with and without an astrophysical x-ray background. The difference between the two curves quantifies the improvement in sensitivity to DM decay that is achieved by modeling the astrophysical X-ray background. The systematic uncertainty arising from the choice of background model is already incorporate into our analysis. Including the background model improves the lower bound on DM decay lifetime by approximately two orders of magnitude for DM masses below 100~MeV, where the DM signal overlaps significantly with the astrophysical components. 

At higher masses, the improvement that stems from including an astrophysical background model decreases significantly. This is due to the fact that the spectrum of the astrophysical background model, which is primarily due to inverse-Compton scattering at high energies, begins to look similar to the spectrum produced by high mass DM. Since our astrophysical background model is conservative, and does not attach a penalty to decreasing the amplitude of the astrophysical background model (so long as the fit to the data continues to be good), our standard model produces a very conservative dark matter limit that is similar to the ``no-background" limit. This illustrates that any systematic uncertainties in our astrophysical background model do not significantly affect the strength of our analysis, because they only serve to improve the limits in scenarios where DM cannot match the spectrum of the x-ray data on its own. Since the background model primarily improves constraints below the QCD scale, where only the $e^{+}e^{-}$ channel is available, our X-ray constraints on dark photon and scalar DM particle above 100~MeV remain unaffected.

Finally, we assess the theoretical uncertainty arising from the choice of DM density profile, which enters our analysis through the $D$-factor. While the main results are presented for the standard NFW profile, here we examine how the constraints change under alternative profile assumptions. Specifically, we recompute the total flux for the Einasto~\cite{Navarro:2008kc}, Burkert~\cite{Burkert:1995yz}, and Moore~\cite{Diemand:2004wh} profiles, using the parameters in Ref.~\cite{Cirelli:2010xx}, and compare them to the NFW baseline.

In Fig.~\ref{fig:NoBgk} we show the resulting envelope (gray bands) of lifetime constraints across these profiles. The bounds can shift downward by approximately 20\% in the case of the flatter Burkert profile, and upwards by approximately 10\% in the case of the steeper Moore profile. We emphasize, however, that decay limits scale only linearly with the $D$-factor. Unlike annihilation constraints, our results are insensitive to the inner slope of the DM profile. As a result, the profile dependence remains a subdominant source of uncertainty in our analysis. In particular, at low masses ($m_{\chi}\leq 100$~MeV) the systematic uncertainty from modeling the astrophysical background, which is reflected in the gap between the teal and purple curves, exceeds the profile uncertainty by roughly an order of magnitude, and therefore dominates the overall error budget.

\bibliography{main}

\end{document}